# Polarization-Encrypted Orbital Angular Momentum Multiplexed Metasurface Holography

Hongqiang Zhou,[†] Basudeb Sain,[‡] Yongtian Wang,[*,†] Christian Schlickriede,[‡] Ruizhe Zhao,[†] Xue Zhang,[†] Qunshuo Wei,[†] Xiaowei Li,[§] Lingling Huang,[*,†] Thomas Zentgraf[*,‡]

[†]School of Optics and Photonics, Beijing Institute of Technology, Beijing, 100081, China

[‡]Department of Physics, Paderborn University, Warburger Straße 100, 33098 Paderborn, Germany

[§]Laser Micro/Nano-Fabrication Laboratory, School of Mechanical Engineering, Beijing Institute of Technology, Beijing 100081, China

[*]Email: wyt@bit.edu.cn
[*]Email: huanglingling@bit.edu.cn
[*]Email: thomas.zentgraf@uni-paderborn.de

ABSTRACT Metasurface holography has the advantage of realizing complex wavefront modulation by thin layers together with the progressive technique of computer-generated holographic imaging. Despite the well-known light parameters, like amplitude, phase, polarization and frequency, the orbital angular momentum (OAM) of a beam can be regarded as another degree of freedom. Here, we propose and demonstrate orbital angular momentum multiplexing at different polarization channels using a birefringent metasurface for holographic encryption. The OAM selective holographic information can only be reconstructed with the exact topological charge and a specific polarization state. By using an incident beam with different topological charges as erasers, we mimic a super-resolution case for the reconstructed image, in analogy to the well-known STED technique in microscopy. The combination of multiple polarization channels together with the orbital angular momentum selectivity provides a higher security level for holographic encryption. Such a technique can be applied for beam shaping, optical camouflage, data storage, and dynamic displays.

Through encapsulating abrupt phase discontinuities at patterned interfaces, namely metasurface, a versatile field of planar optics has been developed, which provides a paradigm for shifting the wavefront modulation from bulk elements to ultrathin surfaces.[1,2] Metasurfaces can provide versatile platforms to tailor electromagnetic wavefront by designing two-dimensional arrays of nanoscale antennas or resonators.[3-10] Metasurfaces have successfully provided a wide range of practical applications in miniaturized optics, including beam shaping, optical cloaking, holographic displays, nonlinear optics, optical encryption, and data storage.[11-20] Especially, metasurface holography can conquer the challenges faced by traditional holography, and possess the advantages of a large field of view, elimination of unwanted diffraction orders and high information capacity. By exploring the phase, amplitude and polarization design flexibilities of metasurfaces, such holographic technique can achieve versatile multiplexing and information processing capabilities, surpassing their traditional counterparts.[21-24]

As one of the fundamental physical properties, orbital angular momentum (OAM) has attracted much attention. Vortex beams that carry OAM modes are characterized by a donut-shaped intensity distribution and exhibit helical phase factors $\exp(il\varphi)$, where $l$ indicates the topological charge number and $\varphi$ is the azimuthal angle.[25] OAM is considered to be the last efficient dimension of optical multiplexing. Due to the orthogonality between the different OAM modes, they can be superimposed to enhance the information capacity. Therefore, OAM modes are suitable to realize large-capacity optical communication systems, optical tweezers and so on.[26-32] Besides, OAM has also found many applications in quantum information processing, two/multi-photon entanglement and so on.[29,33-42] Using the properties of orthogonality between different OAM modes and the OAM conservation law, OAM selective holography has the potential to achieve optical data storage with complex security features.[43-45] However, the major problems are the sparse sampling and the requirement of post-processing to eliminate crosstalk between the images.

Here, we propose and demonstrate OAM and polarization selective holography for information encryption and image generation using all-dielectric birefringent metasurfaces. The polarization selectivity of such metasurface relies on the birefringent response to the incident light. The OAM states with different topological charges provide extra security and design freedom for holographic encryption compared to previously reported metasurface holography schemes. For a vortex beam carrying a specific OAM, together with a suitable combination of input and output polarization states, different holographic images can be reconstructed at the detection plane. Ideally, there is no crosstalk between the different output polarization channels. Besides, we realize a vortex array in the orthogonal polarization channel as the indicator of the angular momentum carried out by the input beam. Furthermore, such OAM selective holography can provide extra flexibility for erasing and modifying holographic reconstructed images. By utilizing OAM beams to depict particular image detail while preserving useful aspects of the images, some hidden information (camouflage) can be revealed, providing a more sophisticated scene than the plane wave illumination case. Such a concept can mimic the stimulated emission depletion (STED) method in regard to the use of OAM beams as erasers. To illustrate this kind of mimicking effect, we show experimental results for different kinds of illumination conditions. A larger OAM mode selectivity for optical holography may provide higher flexibility in optical encryption, larger data storage capacity, and innovation in dynamic displays.

RESULTS AND DISCUSSION

In the following, we demonstrate the design of multiple OAM modes and holograms, as shown in Figure 1. When illuminating the metasurface with a Gaussian beam, no image information is observed in the $T_{xx}$ and $T_{yy}$ channels (subscript $ab$ denotes the input and output polarization channels), while an array of vortex beams with spatially variant topological charges can be observed in the cross-polarization $T_{xy}$ and $T_{yx}$ channels. By illuminating the same metasurface with a vortex beam carrying the exact topological charge as designed, the target hologram can be reconstructed due to the OAM selectivity. For example, the word "NATURE" or "SCIENCE" shows up in the co-polarization channels (with topological charge $l$=40 or $l$=20, respectively), while the vortex array changes its spatial distributions accordingly in the cross-polarization channel ($T_{xy}$ and $T_{yx}$), serving as an indicator for the incident OAM. Such a reconstruction scheme might be counterintuitive at first sight. However, we purposely design the holograms by integrating the OAM phase distributions. Note, the original image is sampled to dot arrays according to the diameter sizes of different vortex beams (detailed explanation can be found in the Supporting Information). Such a sampling method ensures that the vortex beams with the accurate OAM state can generate the desired holographic pattern by obeying the angular momentum conservation law, while vortex beams with different OAM cannot generate the high-resolution images due to the spatial overlap, implying different vortex beams will only reconstruct strong blurred images.

For the design of multiple polarization channels, we utilize a birefringent dielectric metasurface composed of nanofin arrays with different rectangular cross-sections. Each nanofin can be regarded as a Jones matrix, which connects the input field to the output field. Due to the non-chiral symmetry of the meta-atom, the anti-diagonal matrix elements $T_{xy}$ and $T_{yx}$ are identical. Note that the near-field coupling between neighboring nanofins is negligible. Such unitary Jones matrix can be decomposed into the multiplication of two rotation matrices and a diagonal matrix as follows:[46]

$$J = V \begin{bmatrix} e^{i\varphi_{xx}} & 0 \\ 0 & e^{i\varphi_{yy}} \end{bmatrix} V^T = R(\theta) \Delta R(-\theta) \quad (1)$$

Where $\Delta$ indicates the eigenvector of the Jones matrix $J$ and R represents a rotation matrix with the orientation angle $\theta$. The $\varphi_{xx}$, $\varphi_{yy}$ are polarization-dependent phase shifts.

For multiplexing in all the four linear polarization channels, we purposely design the Jones matrix with the transmission amplitude $A$ of nanofin as follows:

$$J = A \begin{bmatrix} e^{i(\varphi_{xx} - l_1\varphi)} & e^{i(\varphi_{yx} - l_2\varphi)} \\ e^{i(\varphi_{yx} - l_2\varphi)} & e^{i(2\varphi_{yx} - \varphi_{xx} + \pi - l_3\varphi)} \end{bmatrix} \quad (2)$$

Hence, we acquire another phase modulation freedom in the cross-polarization channel $\varphi_{yy} - l_3\varphi = 2(\varphi_{yx} - l_2\varphi) - (\varphi_{xx} - l_1\varphi) + \pi$. Note that the characteristic phase distributions of $\exp(il\varphi)$ are included to obtain the OAM selectivity and that different topological charges $l$ can be independently chosen in the different polarization channels. As mentioned above, when the meta-holograms are illuminated with polarized light carrying the opposite OAM, the corresponding encoded phases ($l_1$, $l_2$ or $l_3$) are "quenched" and the output light field possess the required phase of $\varphi_{xx}$, $\varphi_{xy}$ (or $\varphi_{yx}$) and $\varphi_{yy}$. While on the other hand for unmatched OAM values the reconstructed images become blurry. We use a smart hologram design algorithm so that the phase modulations in the three different polarization combinations become independent of each other with no crosstalk between them. While for the camouflage case, mimicking STED as demonstrated in the following, we utilize both phases, with and without the OAM modes of $\exp(il\varphi)$, for encoding two similar images (one image with extra details compared to the other one) within one channel. Hence, by illuminating either by the plane wave or the vortex beam, vivid holographic images can be reconstructed. However, only the incident light carrying the designed OAM can reveal some sophisticated details. The details of the developed design algorithm can be found in the Supporting Information.

We experimentally demonstrate the two cases of OAM selective hologram and camouflage. The scanning electron microscope (SEM) images of the fabricated metasurfaces composed of nanofins are as shown in Figure 2a-d. Simulation and fabrication details can be found in the Methods section. The results for the OAM selective hologram in the co-polarization channels and OAM indicators in the cross-polarization channels are shown in Figure 3. For an incident plane wave with $l=0$, the reconstructed image information is hidden in the $T_{xx}$, $T_{yy}$ channels (Figure 3a-d), while for the illuminating light carrying the specific OAM ($l$=40 with $x$ polarization or $l$=20 with $y$ polarization illumination) the desired images of "NATURE" and "SCIENCE" are reconstructed in the $T_{xx}$, $T_{yy}$ channels, respectively (Figure 3e-h). Note that in the experiment, a beam spot (residual light) with donut shape appears as a zero-order diffraction spot due to the discrete nature of the metasurface and the deviation of the sample fabrication from the theoretical design. The peak signal to noise ratios (PSNRs) of the images 'NATURE' and 'SCIENCE' in Figure 3g,h are 29.11 dB and 29.83 dB, respectively. While in the cross-polarization channels ($T_{xy}$ and $T_{yx}$), a vortex array with different radii of the donut is generated, serving as an indicator of the input angular momentum. Such spatially variant vortex array carries the OAM related to the diffraction orders, which can be expressed as $l=m \times l_x$. Here $m$ is the diffraction order (0, ±1, ±2, ...), and $l_x$ is the predefined value (we set $l_x$=20). Note that we purposely optimize only five diffraction orders of vortices while suppressing all the other unwanted orders in the phase design. According to the angular momentum conservation law, when the input beam carrying a topological charge of $b$, the vortex array obeys $l=m \times l_x+b$. Hence, the corresponding vortex beam can be annihilated when its topological charge is opposite to the input beam, generating a bright central spot. By observing the position of the brightest spot, one can detect the topological charge of the incident OAM. Depending on the input beam (plane wave or vortex beam), the vortex array behaves differently in the orthogonal polarization channels with the topological charges $l$=0, 10, 20, 40, as shown in Figure 3i-p. As a step of pre-processing, we customize the sampling of the original images to the radius of the specific OAM. Therefore, illuminating the sample with other OAM beams results in the overlap of lower intensity donuts that cause blurry in the image and the clear image formation is hidden.

Furthermore, we design another metasurface hologram sample by mimicking the STED concept with regard to sub-diffraction resolutions. Such a STED-like method utilizes vortex beams as an 'eraser' to suppress/deplete a portion of the excited fluorescence and increase the image resolution. We utilize this analogy to reveal additional image features with higher-order OAM beams. We choose a bunch of grapes and a leaf for both Gaussian beam and vortex beam reconstruction in the co-polarization channels $T_{xx}$ and $T_{yy}$ (Figure 4). For the Gaussian beam illumination, the $T_{xx}$ channel reconstructs the image of the grapes (as shown in Figure 4a,c). When the incident light carries a topological charge of $l$=40 with $x$ polarization, the grapes show some additional detailed textures (Figure 4b,d). In addition, the $T_{yy}$ polarization channel shows the reconstruction of a filled leaf for the Gaussian beam illumination (Figure 4e,g). Again, when $l$=40 is carried by the incident beam, a sawtooth-shape appears at the edge of the filled leaf, which demonstrates a serrated leaf in Figure 4f,h. In this case, the OAM input beam works like an eraser, which can realize an effect like image depletion and therefore a reshaping of the image information (*e.g.,* revealing more details of the image). Note that the zero-order spots are still observable at the center of the holographic images due to fabrication inaccuracies as well as the discrete nature of the metasurface. The lower contrast for the vortex beam illumination compared to that of the plane wave case is due to the fact that the vortex beam has a donut-shaped intensity distribution. Hence the efficiency of the metasurface is lower. Using the advantage of the OAM selectivity and the polarization channel combinations, including the circular polarization channels, several possibilities for applications exist in encryption and decryption of information (see Supporting Information).

CONCLUSIONS

In summary, we propose and demonstrate several holographic encryption and display strategies using orbital angular momentum and multiple polarization selective channels. OAM selective holography together with optical OAM indicators, and optical camouflage by depicting sophisticated detailed information are realized within a single metasurface by choosing different input/output polarization channels. With the help of the orthogonality of multiple OAM states, any possible OAM combination can be optimized for the same polarization channel, whereas the incident OAM beam works like a key to unlock the specific information. Such a method further enhances the information capacity and enriches the potential applications in encryption and display technologies.

METHODS

**Simulation**. We adopted dielectric silicon nanofins as building blocks for the metasurface holograms. For optimizing the optical properties, we fix the height and the period of the silicon nanofins at 600 nm and 400 nm, respectively, and sweep the length and width in the range of 80–280 nm for a working wavelength of 800 nm. The 2D parameter optimization is done using a rigorous coupled-wave analysis (RCWA) method.

**Fabrication**. The all-silicon metasurfaces are fabricated on a $SiO_2$ substrate. First, an amorphous silicon (a-Si) film was deposited to 600 nm by plasma-enhanced chemical vapor deposition (PECVD). A poly-methyl-methacrylate resist layer was spin-coated onto the a-Si film. To remove the solvent, it is baked on a hot plate at 180 °C for 2 min. Then, standard electron beam lithography is used to pattern the desired structure and subsequent development in 1:3 MIBK:IPA solution. Next, the sample was washed with IPA before being coated with a 45-nm-thick chromium layer by electron beam evaporation. Next, we performed the lift-off process in hot acetone. In the last step, the desired structure was transferred from chromium to silicon by using inductively coupled plasma reactive ion etching (ICP-RIE).

ACKNOWLEDGMENTS

The authors acknowledge the funding provided by the National Key R&D Program of China (No. 2017YFB1002900) and the European Research Council (ERC) under the European Union's Horizon 2020 research and innovation program (grant agreement No. 724306). We also thank the NSFC-DFG joint program (DFG No. ZE953/11-1, NSFC No. 61861136010) for additional support. L.H. acknowledge the support from Beijing Outstanding Young Scientist Program (BJJWZYJH01201910007022), National Natural Science Foundation of China (No. 61775019) program, Beijing Municipal Natural Science Foundation (No. 4172057), Beijing Nova Program (No. Z171100001117047) and Fok Ying-Tong Education Foundation of China (No.161009). We also thank Dr. Xin Li for the discussion of the algorithm.

ASSOCIATED CONTENT

**Supporting Information**

The Supporting Information mainly includes the algorithm flowchart, the metasurface design, and extra experimental verification of our design. This material is available free of charge *via* the Internet at http://pubs.acs.org.

AUTHOR INFORMATION

**Author Contributions**

Y.W., L.H., and T.Z. proposed the idea, H.Z., R.Z., and X.Z. conducted the pattern designs and numerical simulations, H.Z. conducted the hologram generations, B.S. fabricated the samples, H.Z, B.S., and C.S. performed the measurements, L.H., H.Z., B.S., and T.Z. prepared the manuscript. Y.W., L.H., and T.Z. supervised the overall projects. All the authors analyzed the data and discussed the results.

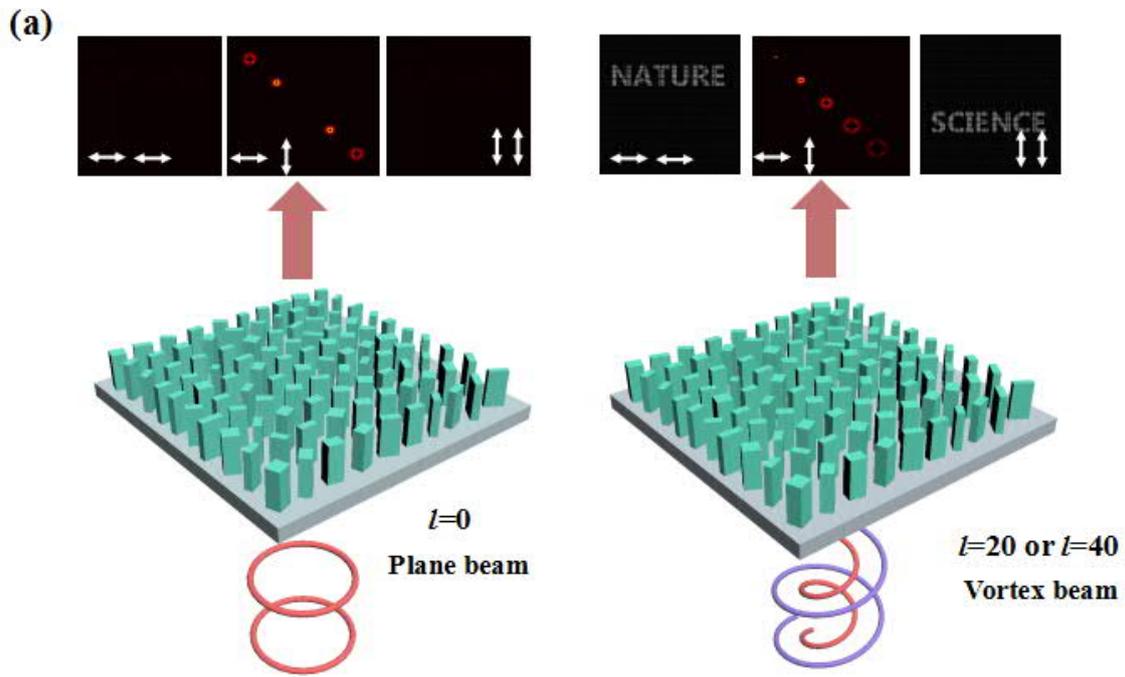

**Figure 1.** Vortex and polarization encryption principle diagram. For the illumination of the metasurface hologram by either Gaussian beams ($l=0$) or vortex beams carrying a specific OAM ($l=20$ or $40$), distinct images are reconstructed for different combinations of the input and output polarizations. The white arrows show the incident (first) and image (second) polarization state.

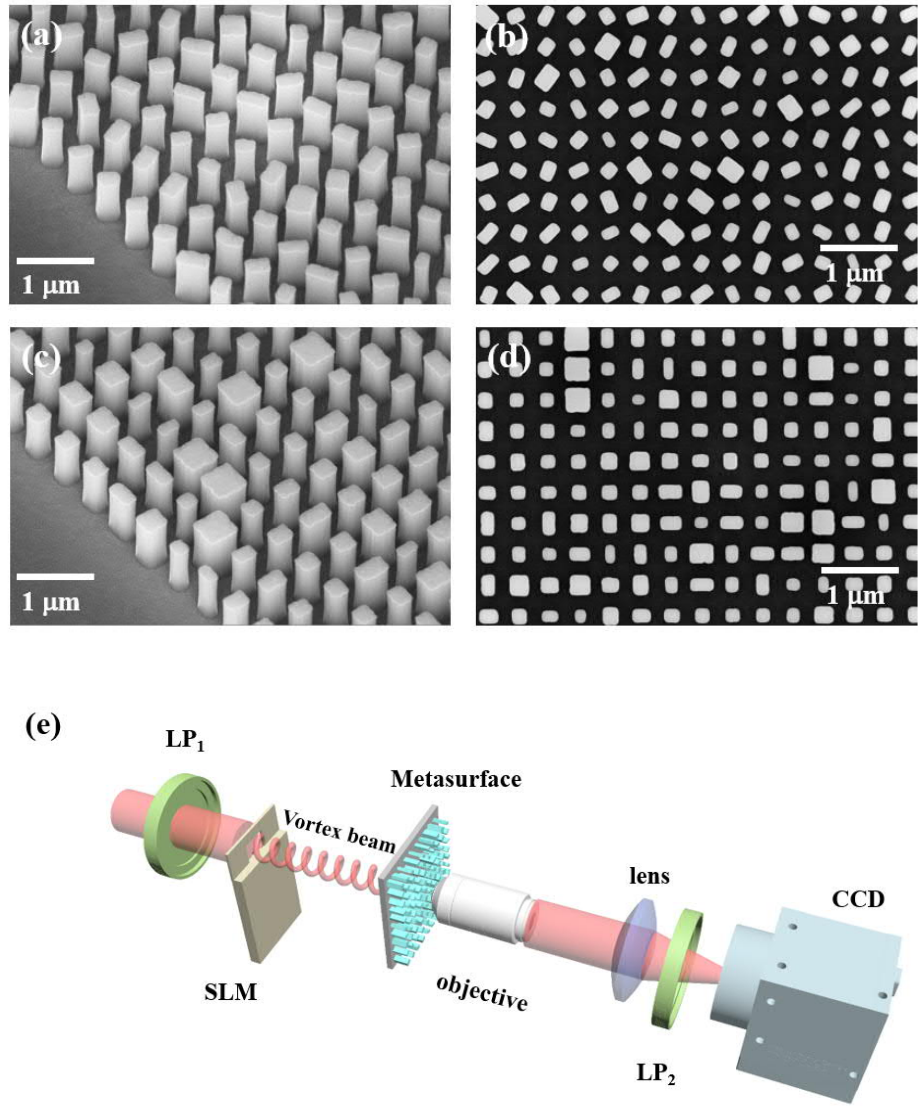

**Figure 2.** Fabricated samples and optical setup. (a)–(d) Scanning electron microscopy images of the fabricated metasurface samples for a top and 45° side view. Scale bar corresponds to 1 μm. (e) Schematic illustration of the optical setup for characterizing the holograms. A spatial light modulator (SLM) was used for generating the vortex beams. Two linear polarizers (LP$_1$ and LP$_2$) are used for setting and detecting the polarization state.

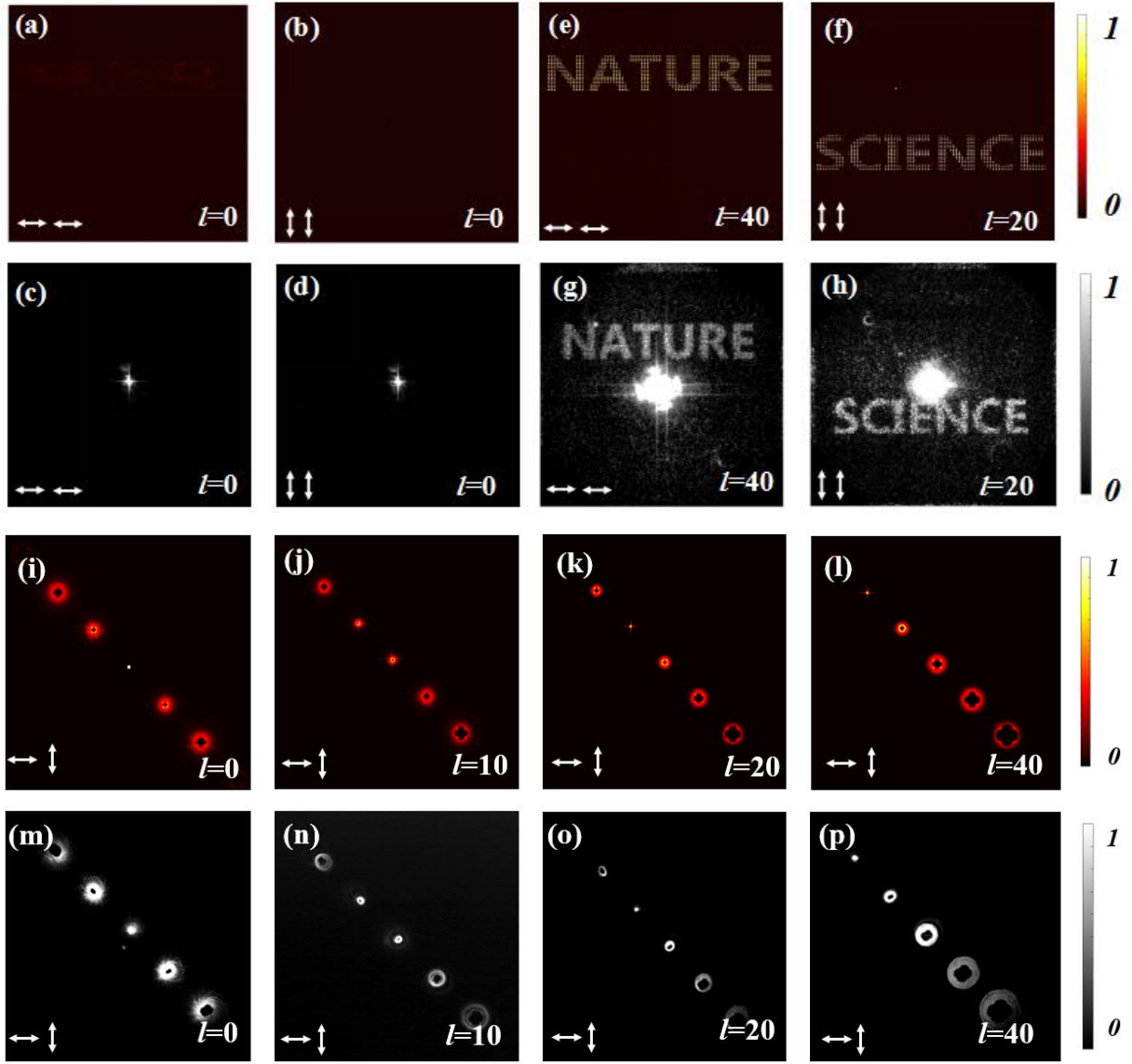

**Figure 3.** OAM indicator and OAM holography in different polarization channels. (a-b) $T_{xx}/T_{yy}$ simulation results for illumination with $l=0$. (c-d) Experimental reconstruction ($T_{xx}/T_{yy}$) with $l=0$. Only the zero-order spot is observable in the Fourier plane. (e-f) $T_{xx}/T_{yy}$ simulation results for the same metasurface with $l=40,20$. The both channels show the words "NATURE" and "SCIENCE" depending on the polarization and the OAM. (g-h) Experimental reconstruction ($T_{xx}/T_{yy}$) with $l=40,20$. (i-l) Simulation results for the illumination with $l=0,10,20,40$ and the cross-polarization channel ($T_{yx}$). (m-p) Experimental reconstruction with $l=0,10,20,40$ in the cross-polarization channel ($T_{yx}$).

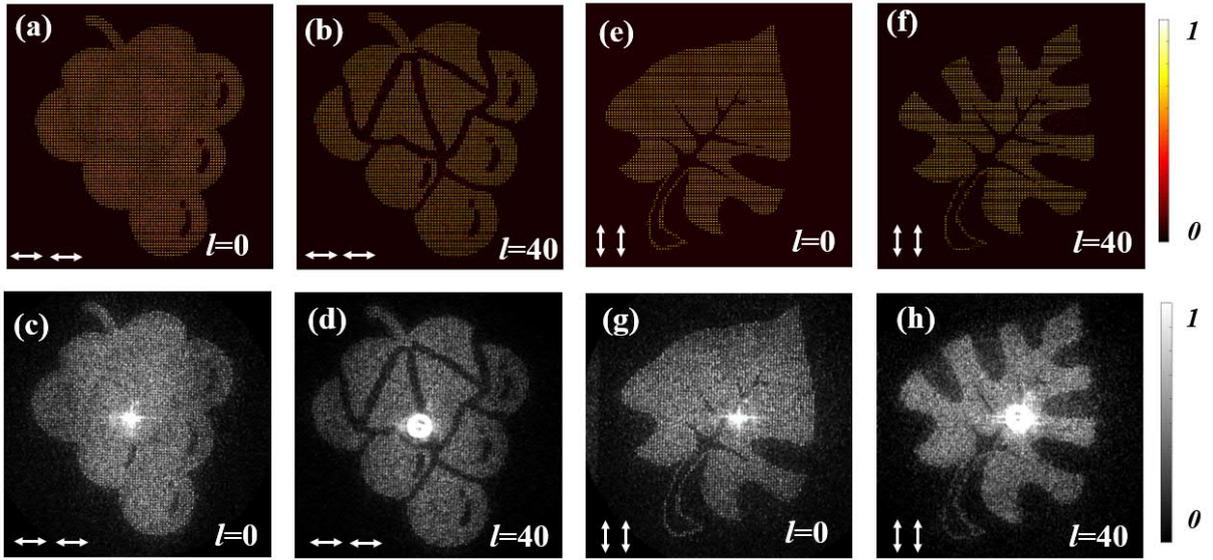

**Figure 4.** Holographic camouflage with plane wave and vortex beam illumination mimicking the STED-like method. (a-b) $T_{xx}$ simulation with $l=0$ and $l=40$. (c-d) $T_{xx}$ experimental reconstruction with $l=0$ and $l=40$. (e-f) $T_{yy}$ simulation with $l=0$ and $l=40$. (g-h) $T_{yy}$ experimental reconstruction with $l=0$ and $l=40$.